\documentstyle[12pt]{article}
\begin{document}

\title{Linear $r$--Matrix Algebra for a Hierarchy of
One--Dimensional Particle Systems Separable in Parabolic Coordinates}
\author{J C Eilbeck${}\sp 1$, V Z Enol'skii${}\sp {1,2}$,\\
V B Kuznetsov${}\sp 3$, D V Leykin${}\sp2$\\
${}\sp 1$ Department of Mathematics, Heriot-Watt University\\
Riccarton, Edinburgh EH14 4AS, Scotland\\
${}\sp 2$ Department of Theoretical Physics\\
Institute of Metal Physics \\
Vernadsky str. 36, Kiev-680, 252142, Ukraine \\
${}\sp3$Department of Mathematical and Computational Physics\\
Institute for Physics, University of St Petersburg\\
St Petersburg 198904, Russia}
\renewcommand{\theequation}{\arabic{section}.\arabic{equation}}
\newcommand{\PB}{\stackrel{\textstyle\otimes}{,}}
\newtheorem{theorem}{Theorem}
\newtheorem{prop}{Proposition}
\maketitle

\begin{abstract} We consider a hierarchy of many-particle systems on the
line  with polynomial potentials separable in parabolic coordinates.
The first non-trivial member of this hierarchy is a generalization
of an integrable case of the H\'enon-Heiles system.  We give a  Lax
representation in terms of $2\times 2$ matrices for the whole hierarchy
and construct  the associated linear $r$-matrix algebra with the
$r$-matrix dependent on the dynamical variables. A Yang-Baxter equation
of dynamical type  is  proposed.  Classical integration in a particular
case is carried out and quantization  of the system is discussed  with
the help of separation variables.  This paper was published in the rary
issues:  Sfb 288 Preprint No. 110, Berlin and {\em Nonlinear Mathematical
Physics}, {\bf 1(3)}, 275-294 (1994) \end{abstract}

\section{Introduction}
\setcounter{equation}{0}

In the last decade, much work has been carried out in the study of
completely integrable systems  admitting  a  classical  $r$-matrix
Poisson structure (c.f. \cite{fa84,ft87}) where the $r$-matrices depend
{\it only} on the spectral parameters.  More recently, interest has
developed in the study of completely integrable systems  where the
$r$-matrices depend also on dynamical variables
\cite{bv90,fm91,ma85,se83}.  It is remarkable that the celebrated
Calogero--Moser system, whose complete integrability  was  shown  a
number of years ago (see e.g. \cite{op81}), has been found only
recently to  possess a classical $r$--matrix of this dynamical type
\cite{at92}. The development of this ideology came to the series of
interesting papers   \cite{skl93},\cite{bs93},\cite{bb93},\cite{bab93},\cite{bab94},\cite{br94}.

In this paper we study another example of a dynamical
$r$--matrix structure. It is already well known that it is possible to provide $2\times 2$ Lax
operator satisfying the standard linear $r$-matrix algebra for the
Hamiltonian system of natural form where the potential ${\cal U}$ is of
second degree by the coordinates (see e.g.
\cite{se83,ga83,sk89,ku92,kuz92}). We consider a  hierarchy of one--dimensional
many-particle systems  with  polynomial  potentials of higher degree whose complete
integrability in the framework of the Lax representation  is  known,
but for which the associated $r$--matrix  Poisson  structure has  not
been discussed.  The systems represent a generalization  of  the
known  hierarchy  of two-particles  systems with  polynomial
potentials  separable  in parabolic coordinates, which have  the
form   (see,  e.g.  \cite{hi87,pe91}).
\begin{equation}
{\cal V}_N=\sum_{k=0}^{[N/2]}
 2^{1-2k}\left({}^{N-k}_{\quad k}\right) q_1^{2k}
q_2^{N-2k},\label{pot0}
\end{equation}
where the positive integer $N$ enumerates the members of the hierarchy.  We study an
integrable many particle generalizations of (\ref{pot0}) in  the
framework of  the  Lax  representation,  written in  terms  of  $2\times  2$
matrices,
\begin{eqnarray}
\dot L^{(N)}(z)&=&[M_{N}(z),L_{N}(z)],\nonumber\\
L_{N}(z)&=&\left(\begin{array}{cc} V(z)&U(z)\\
W_{N}(z)&-V(z)\end{array}\right),\quad
M^{(N)}(z)=\left(\begin{array}{cc} 0&1\\
Q_{N}(z)&0\end{array}\right),\label{*} \end{eqnarray} where the
functions $U(z),\,V(z),\,W_N(z),\,Q_N(z)$ depend rationally in the
spectral parameter $z$ and some constraints are  imposed on  them.
In  the first nontrivial case ($N=3$) the system can be considered as
a many-particle generalization of a known integrable case of the
H\'enon-Heiles system \cite{hh64}. We emphasize, that the ansatz
(\ref{*}) is a natural generalization of the Lax representation found
in \cite{ntz87} to describe an integrable case of the H\'enon-Heiles
system. See also \cite{ku92},\cite{kuz92} for the link to the
$su(1,1)$--Gaudin magnet which corresponds to a free $n$-dimensional
particle separable in parabolic coordinates.

The polynomial second order spectral problem associated with the Lax
representation was studied in \cite{zl90} in the framework of a
$(K\times K)$ Lax representation, and the recursion relations between
different members of a hierarchy of Lax matrices was also given, but no
explicit treatment of any associated dynamical system was discussed.
The $(2\times 2)$ Lax representation we describe appears to us to be
more useful to investigate the class of integrable systems under
consideration and to elucidate the classical $r$-matrix Poisson
structure of the system. Such Lax representations were introduced in
\cite{wo85},\cite{wo86} and their connection with the restricted
coupled KdV flows was studied in \cite{rw91},\cite{ar92},\cite{arw92} (see also \cite{ntz87} for
the isomorphism with KdV). But the $r$-matrix Poisson structure
asosiated with higher degree potentials has not been discussed.

We consider the system  within the method of variable separation
\cite{kmw76,sk89,ku92} that permits us to develop the classical
theta-functional integration theory and to consider the associated
quantum problem.  The last problem is reduced to a set of
multiparameter spectral problems which are a confluent form of ordinary
differential equations of the Fuchsian type.

The central result of the paper is the description of the Poisson
structure of the system by a dynamical linear $r$--matrix algebra
defined in $V^2\otimes V^2$ \cite{bv90,ma85,se83},
\begin{equation}
\{L_1^{(N)}(x)\PB L_2^{(N)}(y)\}=[d^{(N)}_{12}(x,y),L_1^{(N)}(x)]-
[d^{(N)}_{21}(x,y)L_2^{(N)}(y)],\label{rsal}
\end{equation}
where $\PB$ is the direct product $\otimes$ of two matrices, but with
the product of two matrix elements replaced by their Poisson bracket.
The $L$ matrices are defined in  $V^2\otimes V^2$ by $L_1^{(N)}(x)=
L^{(N)}(x)\otimes I,\,L_2^{(N)}(y) = I \otimes L^{(N)}(y)$,  $I$ is the
$2\times2$ unit matrix, $d^{(N)}_{12}$ and $d^{(N)}_{21}$ are matrices
depending both on  spectral parameters and on dynamical variables
through the factor $(Q_N(x)-Q_N(y))/(x-y)$.

In the space $V^2\otimes V^2\otimes V^2$, the matrices $d^{(N)}_{ij}$ satisfy the
constraints (classical Yang--Baxter equation of dynamical type)
\begin{eqnarray}
&&[d^{(N)}_{12}(x,y),d^{(N)}_{13}(x,z)]+[d^{(N)}_{12}(x,y),d^{(N)}_{23}(y,z)]
+[d^{(N)}_{32}(z,y),d^{(N)}_{13}(x,z)]+\nonumber\\
&&+\{L_2^{(N)}(y)\PB d^{(N)}_{13}(x,z)\}+\{L_3^{(N)}(z)\PB
d^{(N)}_{12}(x,y)\} +\nonumber\\
&&+[c^{(N)}(x,y,z),L_2^{(N)}(y)-L_3^{(N)}(z)]=0\label{ybe}
\end{eqnarray}
plus  cyclic permutations. The matrices $d^{(N)}_{ij}$ and $c$  in
(\ref{ybe}) involve the dynamical variables through the factor $Q_N$.
We follow the now standard but somewhat confusing notation that
$L_i^{(N)}, d^{(N)}_{ij}$ refer to different matrices depend on whether
the current space is  $V^2\otimes V^2$ or  $V^2\otimes V^2\otimes V^2$.
In $V^2\otimes V^2\otimes V^2$, $L_1^{(N)}(x)= L^{(N)}(x)\otimes I\otimes
id,\,L_2^{(N)}= I  \otimes L^{(N)}(y)\otimes I,\, L_3^{(N)}(z)=L^{(N)}(z)\otimes
I\otimes I$.  The matrix $d^{(N)}_{ij}$ in $V^2\otimes V^2\otimes V^2$ acts like
$d^{(N)}_{ij}$ in $i$th and $j$th space and as $I$ in the third space.
Except for the final term, eqn. (\ref{ybe}) is the same as that given in
\cite{ma85}. The final term in the dynamical Yanng-Baxter equation presents as it is emphasised in \cite{skl93} some new matrix, for which a more general ansatz is given in \cite{skl93}. 

A main theme of the paper is to present
a new solution of the dynamical Yang-Baxter equations
(\ref{ybe}) which is associated with a  hierarchy  of
one-dimensional dynamical systems. At the same time the $r$--matrix
structure of parabolic coordinates is elucidated.

The main results were announced in the paper \cite{eekl93aa}. The
classical systems separated in ellipsoidal and spheroidal coordinates
are concidered in \cite{eekt94}.

\section{The Hierarchy of Separable Systems}
\setcounter{equation}{0}
In this section we describe a hierarchy of completely integrable
one-dimensional many-particle systems with  polynomial potentials.
The different members of the hierarchy are connected by  recurrent
relations.  We give the Lax representation in terms of $(2\times 2)$
matrices for all  the hierarchy, describe the associated algebraic
curves and present explicit formulae for the integrals of motion.

\subsection{Hierarchy of Hamiltonian Systems with Polynomial Potentials}
Let us consider the hierarchy of
the Hamiltonian systems of $n+1$ particles defined by the Hamiltonians
\begin{equation}
H_N(p_1,\ldots,p_{n+1};q_1\ldots,q_{n+1})={1\over
2}\sum_{i=1}^{n+1} p_i^2 +{\cal U}_N (q_1,\ldots,q_{n+1}),\label{ham}
\end{equation}
where the potentials ${\cal U}_N$  for each  member of the hierarchy
are given by the
recurrence relation
\begin{eqnarray}
{\cal U}_{N}&=&(q_{n+1}-B){\cal U}_{N-1}+{1\over 4}
\sum_{i=1}^n\sum_{j=2}^N (-1)^{j}q_i^2{\cal
U}_{N-j}A_i^{j-2}\label{hipot}
\end{eqnarray}
with the first trivial potentials given as
\begin{eqnarray}
{\cal U}_0&=&0\nonumber\\{\cal
U}_1&=&-2q_{n+1}-2B,\nonumber\\ {\cal
U}_2&=&-2q_{n+1}^2-{1\over 2}\sum_{i=1}q_i^2.\label{utr}
\end{eqnarray}
At $A_i=0,\,i=1,\ldots,n,\,B=0$ the recurrent relations (\ref{hipot})
can be solved as
\begin{eqnarray}
&&{\cal V}_N(q_1,\ldots,q_{n+1})= -\sum_{k=0}^{[N/2]}
 2^{1-2k}\left({}^{N-k}_{{\quad}k}\right) \left(\sum_{i=1}^n
q_i^2\right)^k q_{n+1}^{N-2k}\label{pot1}
\end{eqnarray}
yielding the principal term of ${\cal U}_N(q_1,\ldots,q_{n+1})$ which
is a many-particle generalization of (\ref{pot0}). The lower degree terms
are introduced so as to make the many-particle generalization non-degenerate. The  expression for the
potentials ${\cal U}_N$ at $N>2$ can be written in the form of
$(N-2)\times(N-2)$ determinant
\begin{equation}
{\cal U}_N=(-1)^{N-1}\,{\rm
det}\,\left(\begin{array}{cccccc} f_N&g_{-1}&g_0&g_1&\ldots&g_{N-5}\\
f_{N-1}&-1&g_{-1}&g_0&\ldots&g_{N-6}\\
f_{N-2}&0&-1&g_{-1}&\ldots&g_{N-7}\\
\vdots&{}&\ddots&\ddots&\ddots&\vdots\\
f_4&0&\ldots&0&-1&g_{-1}\\
f_3&0&\ldots&\ldots&0&-1\end{array}\right),\label{genpot}
\end{equation}
where
\begin{eqnarray}
g_{-1}=q_{n+1}-B,\quad  g_m&=&{(-1)^m\over 4}\sum_{i=1}^n
A_i^mq_i^2,\,m=0,\ldots\nonumber\\
f_k&=&\sum_{l=0}^2{\cal U}_lg_{k-l-2},\,k=3,\ldots,N,\label{def}
\end{eqnarray}
and the ${\cal U}_0,\,{\cal U}_1,\,{\cal U}_2$ are given by (\ref{utr}). The
first nontrivial potentials are
\begin{eqnarray}
{\cal U}_3&&=-2q_{n+1}^3- q_{n+1}
\sum_{i=1}^nq_i^2 +{1\over 2}\sum_{i=1}^n
A_iq_i\sp 2 +2Bq_{n+1}^2,\label{u3} \\
{\cal U}_4&&=- {1\over 8}\left(\sum_{i=1}^n q_i^2\right)^2 - {3\over
2}q_{n+1}^2\sum_{i=1}^n q_i^2 -2q_{n+1}^4 +\sum_{i=1}^n
A_iq_i^2\left(q_{n+1}-{1\over
2}A_i\right)\nonumber\\&&+B\left(4q_{n+1}^3-2Bq_{n+1}^2+
q_{n+1}\sum_{j=1}^nq_j^2\right).\label{h4}
\end{eqnarray}

The potential (\ref{u3}) is exactly the many-particle generalization of
a known integrable case of the H\'enon-Heiles system for which $n=1$
(c.f.\ \cite{fo91,ntz87}). Analogously the potential
(\ref{h4}) is a many-particle generalization of a $``(1:12:16)"$  system
known to be separable in parabolic coordinates (see e.g. \cite
{hi87},\cite{pe91})

The system with the potential (\ref{u3}) possesses the following
interesting reductions:  a) at $ q_{n+1}=$ {\it const}, it reduces to
the Neuman system, which describes the motion of a particle on a sphere
in the field of a second order potential, and b) at $
q_{n+1}=\sum_{i=1}^n q_i^2$ it reduces to an anisotropic oscillator in
a fourth order potential (see, for instance \cite{pe91}),
\begin{equation} \ddot q_i - 2\sum_{k=1}^n q_k^2q_i +A_i q_i =0,\,
i=1,\ldots,n.\label{4} \end{equation} The same reduction can be carried
out for other members of the hierarchy.

\subsection{The Lax Representation}
We look at a Lax representation of the form (\ref{*}) with
\begin{eqnarray}
U(z)&=& 4z -4q_{n+1}+ 4B -\sum_{i=1}^n{q_i^2\over z+A_i}, \label{uz}\\
V(z)&=&-{1\over 2}\dot U(z),\nonumber\\
&=&2p_{n+1}+\sum_{i=1}^n{p_iq_i\over z+A_i},\label{vz}\\
W_{N}(z)&=&-{1\over 2}\ddot U(z)+U(z)Q_N(z),\label{wz}
\end{eqnarray}
where $Q_N(z)$ is a polynomial of degree
$N-2$. The ansatz for the function $U(z),
V(z),W(z)$ is a generalization of the corresponding ansatz
constructed by Newell et al.\ \cite{ntz87} to give the Lax
representation for the integrable H\'enon-Heiles system. Here we introduce additional degrees of freedom $n>1$, and consider higher degrees of the polynomial $Q_N$. See also \cite{ku92,kuz92} for the link of such ansatz to the $su(1,1)$--Gaudin magnet which corresponds to a free $n$-dimensional particle separable in parabolic coordinates.

\begin{prop}  The Lax representation (\ref{*}) is valid for all the
hierarchy of Hamiltonian systems (\ref{ham}), (\ref{genpot}) with the
polynomial $Q_N(z)$ and the function $W_N(z)$ given by the formulae
\begin{eqnarray}
Q_N(z)&=&zQ_{N-1}(z)-{1\over2}{\partial {\cal
U}_{N-1}(q_1,\ldots,q_{n+1})\over \partial q_{n+1}} ,\label{qn}\\
W_N(z)&=&W^+_N(z)+W^-(z),\label{w+-}\\
W^+_N(z)&=&zW^+_{N-1}-2{\cal U}_{N-1},\quad N=2,\ldots,\label{w+}\\
W^-(z)&=&\sum_{i=1}^n{p_i^2\over z+A_i},\label{w-}
\end{eqnarray}
where ${\cal U}_{N-1}$ is the potential fixing the $N-1$-th member of the
hierarchy.
\end{prop}

\noindent
{\bf Proof}. The Lax representation (\ref{*}) with the functions
(\ref{uz})-(\ref{wz}) is equivalent to the two equations
\begin{eqnarray}
W_N(z)&=&\dot V(z) +Q_N(z) U(z),\label{eqw}\\
\dot W_N(z) &=&2Q_N(z) V(z).\label{eqw'}
\end{eqnarray}
Substituting (\ref{uz},\ref{vz}) and (\ref{qn}) into (\ref{eqw}) we obtain
after some simplification
\begin{eqnarray}
W_N(z)&=&\sum_{i=1}^n{p_i^2\over z+A_i}+z\left(U(z)Q_{N-1}(z)-2{\partial {\cal
U}_{N-1}\over \partial q_{n+1}}-\sum_{i=1}^n{q_i^2\over z+A_i} {\partial {\cal
U}_{N-1}\over \partial q_{i}}\right)\nonumber\\
&+&z\sum_{i=1}^n{q_i^2\over z+A_i} {\partial {\cal U}_{N-1}\over \partial
q_{i}}-\sum_{i=1}^n{q_i\over z+A_i} {\partial {\cal U}_{N}\over \partial
q_{i}}
-2{\partial {\cal U}_N\over \partial q_{n+1}}\nonumber\\
&+&2(q_{n+1}-B){\partial {\cal U}_{N-1}\over \partial
q_{n+1}}+{1\over2}{\partial {\cal U}_{N-1}\over \partial
q_{n+1}}\sum_{i=1}^n{q_i^2\over z+A_i}.\label{3}
\end{eqnarray}
Imposing the conditions (\ref{w+-}), (\ref{w+}), (\ref{w-}) we arrive at the
following recurrence relations for the potentials ${\cal
U}_N(q_1,\ldots,q_{n+1})$
\begin{eqnarray}
{\partial {\cal U}_N\over \partial
q_i}&=&{1\over2}{\partial {\cal U}_{N-1}\over \partial q_{n+1}} q_i
-A_i {\partial {\cal U}_{N-1}\over \partial
q_i},\,i=1,\ldots,n,\label{equ1}\\ {\partial {\cal U}_N\over \partial
q_{n+1}}&=&{\cal U}_{N-1} +{1\over2}\sum_{i=1}^n q_i {\partial {\cal
U}_{N-1}\over \partial q_i}+(q_{n+1}-B){\partial {\cal U}_{N-1}\over
\partial q_{n+1}},\label{equ2}
\end{eqnarray}
where we use (\ref{utr}) to start the recurrence relations.

To prove that the equation (\ref{eqw'}) is also satisfied, we proceed
by induction.  First assume
that it is valid for $N-1$. Then we have
\begin{equation}
\dot W_N(z)=2zQ_{N-1}(z)V(z)-{\partial {\cal U}_{N-1}\over
\partial q_{n+1}}\left(2p_{n+1}+\sum_{i=1}^n{q_ip_i\over
z+A_i}\right).\label{6}
\end{equation}
The left hand side of (\ref{6}) can be rewritten in the form
\begin{eqnarray}
&&{d\over dt}\left(zW^+_{N-1}(z)-2{\cal U}_{N-1} +\sum{p_i^2\over
z+A_i}\right)\nonumber\\&&=z\dot W_{N-1}+2z\sum_{i=1}^n{p_i\over z+A_i}
{\partial {\cal U}_{N-1}\over \partial q_i}-2p_{n+1}{\partial {\cal
U}_{N-1}\over \partial q_{n+1}}\nonumber\\
&&-2\sum_{i=1}^np_i{\partial {\cal U}_{N-1}\over \partial
q_i}-2\sum_{i=1}^n{p_i\over z+A_i}{\partial {\cal U}_N\over \partial
q_i}.\nonumber
\end{eqnarray}
Using induction and the equality (\ref{equ1}) completes the proof.

We can obtain  explicit formulae for the functions $W_N(z)$ and $Q_N(z)$
\begin{eqnarray}
Q_N(z)&=&z^{N-2}-{1\over 2}\sum_{k=0}^{N-3}{\partial {\cal U}_{N-k-1}\over
\partial q_{n+1}} z^k,\label{sumq}\\
W_N(z)&=&4z^{N-1}-2\sum_{k=0}^{N-2}{\cal U}_{N-k-1} z^k+\sum_{i=1}^n
{p_i^2\over z+A_i}.\label{sumw}\end{eqnarray} It easy to see
that\begin{equation}Q_N(x)={1\over 4}{\partial W_N(x)\over \partial
q_{n+1}}\label{qw}
\end{equation}
For example, for the first nontrivial
cases we have
\begin{eqnarray}
Q_3(z)&=&z+2q_{n+1}\label{Q3}\\
W_3(z)&=&4z^2+4zq_{n+1}+ 4Bz+4q_{n+1}^2+\sum_{k=1}^n q_k^2+\sum_{i=1}^n
{p_i^2 \over z +A_i}\label{W3}
\end{eqnarray}
for the many-particle H\'enon-Heiles system and
\begin{eqnarray}
Q_4(z)&=&
z^2+2zq_{n+1}+3q_{n+1}^2+{1\over 2} \sum_{i=1}^n q_i^2-2B
q_{n+1},\label{Q4}\\
W_4(z)&=&4z^3+4Bz^2+4z^2q_{n+1}+4zq_{n+1}^2+z\sum_{i=1}^n
q_i^2\nonumber\\&+&4q_{n+1}^3+2q_{n+1}\sum_{i=1}^nq_i^2-\sum_{i=1}^n
A_iq_i^2-4Bq_{n+1}^2\nonumber\\&+&\sum_{i=1}^n{p_i^2\over z+A_i}\label{W4}
\end{eqnarray}
for the system with $N=4$.

\subsection{Integrability}
The Lax representation yields the hyperelliptic curve
$C^{(N)}=(w,z)$,
\begin{equation}
{\rm Det}\, (L^{(N)}(z)-w I) = 0
\label{curve}
\end{equation}
generating the integrals of motion
$H_N,F_N^{(i)},   i=1,\ldots,n$.  We have
\begin{equation}
w^2={1\over 2}{\rm Tr}(L^{(N)}(z))^2=V^2(z)+U(z)W(z) \label{curve1}
\end{equation}
>From (\ref{curve1}) and (\ref{uz})-(\ref{wz}) we obtain
\begin{equation}
w^2=16z^{N-2}(z+B)^2+8H_N+\sum_{i=1}^n{F_N^{(i)}\over
z+A_i},\quad N=3,\ldots,\label{ccurve}
\end{equation}
where
\begin{eqnarray}
F_N^{(i)}&=&2q_i^2\sum_{j=1}^{N-1}(-1)^{j-1}A_i^j{\cal
U}_{N-j}+4p_{n+1}p_iq_i-p_i^2(A_i+4q_{n+1}-4B)\nonumber\\
&+&\sum_{k,m=1,\,k\neq m}^{k,m=n}{l_{mk}^2\over A_m-A_k},\,
i=1,\ldots,n\label{FN}
\end{eqnarray}
with $l_{km} = q_k p_m-q_m p_k$.
The coefficients of the curve are the integrals of motion,
so noting that the coefficient of $z^{N-1}$ is $32B$ and the
coefficient of $z^{N-2}$ is $16B^2$ we obtain the expressions for the
first potentials (\ref{utr}).  Equating all the other
coefficients of powers of $z$ to zero, i.e. the coefficients of
$z^j,\,j=N-3,\ldots,1$ we reproduce all the
hierarchy of integrable potentials in the form (\ref{hipot}).

\begin{prop} The integrals $H_N$, $F_N^{(i)}, \,i=1,\ldots,n$
are independent and Poisson commute with respect to the standard
Poisson bracket,
\begin{equation}
\{H_N,F_N^{(i)}\}=0,\,\{F_N^{(j)},F_N^{(k)}\}=0,\,
i,j,k=1,\ldots, n.\label{commute}
\end{equation}
\end{prop}

The independence of
the integrals is evident. But we postpone the proof of (\ref{commute})
until \S 4 where the classical $r$--matrix structure will be
elucidated, making the proof trivial.

\section{$r$--Matrix Representation}
\setcounter{equation}{0}
In this section we describe the Poisson structure  associated with the
Lax representation for the hierarchy under consideration.  It  is
found to be a linear $r$--matrix algebra with the $r$--matrices
dependent on dynamical variables and
satisfying some compatibility conditions which are classical Yang--Baxter
equations of dynamical type.  We also  discuss in this section  a
way to describe the system  within the standard linear
$r$--matrix algebra by embedding into a larger dynamical system.

Through all the section we use the standard notations for the Pauli matrices
\begin{equation}
\sigma_0=I=\left(\matrix{1&0\cr0&1}\right),\,
\sigma_1=\left(\matrix{0&1\cr1&0}\right),  \,
\sigma_2=\left(\matrix{0&-i\cr i&0}\right),    \,
\sigma_3=\left(\matrix{1&0\cr0&-1}\right).
\label{pauli}
\end{equation}
Denote $\sigma_{\pm}=\sigma_1\pm i\sigma_2$. Denote as  $P$  the
permutation matrix
$P={1\over 2}\sum_{k=0}^3\sigma_k\otimes\sigma_k$
acting in the product  of two  spaces
$V^2\otimes V^2$. We denote $P_{12}=P\otimes I$, $ P_{13}=
{1\over 2}\sum_{k=0}^3\sigma_k\otimes I\otimes\sigma_k$ and  $P_{23}=I\otimes
P$ the permutation matrices acting in the product of three spaces
$V^2\otimes  V^2\otimes  V^2$. The notation $m_{ik}$ will be used to denote
an (8$\times$8) matrix acting as a unit matrix in $j$-th space $(i\neq j\neq
k)$. For example, let $S=\sigma_-\otimes \sigma_-$,  then $S_{12}=S\otimes
I,\,S_{23}=I\otimes S,\,S_{13}=\sigma_-\otimes I\otimes\sigma_-$.  In addition we
introduce the notation $r_{i,j}(x_i-x_j)=2P_{i,j}/(x_i-x_j)$, $s^{(N)}_{i,j}(x_i,x_j)
=2\alpha(x_i,x_j)S_{i,j}$, where $\alpha(x_i,x_j)$ is a scalar function
depending on dynamical variables.

\subsection{The Classical Poisson Structure}
The following proposition holds:

\noindent
\begin{prop} The classical Poisson structure for the hierarchy
(\ref{uz},   \ref{vz},   \ref{wz}) is written in the form
\begin{equation}
\{L_1^{(N)}(x),L_2^{(N)}(y)\}=[r(x-y),L_1^{(N)}(x)+L_2^{(N)}(y)]+[s(x,y),L_1^{(N)}(x)-L_2^{(N)}(y)],
\label{rsal1}
\end{equation}
where $L_1^{(N)}(x)= I  \otimes
L^{(N)}(x),\,L_2^{(N)}=L^{(N)}(x)\otimes I$, the matrices $r(x-y)$ and $s_N(x,y)$ are
given by the formulae
\begin{eqnarray}
r(x-y) &=&{2\over x-y}P,\quad P= \left(\begin{array}{llll}
1&0&0&0\\
0&0&1&0\\0&1&0&0\\0&0&0&1\end{array}\right)\nonumber\\
&&\label{rmat}\\
s_N(x,y)&=&2\alpha_N(x,y)S,\quad S=\sigma_-\otimes\sigma_-.\label{smat}
\end{eqnarray}
with
\begin{eqnarray}
\alpha_N(x,y)&=&{Q_N(x)-Q_N(y)\over x-y}={x^{N-1}-y^{N-1}\over x-y}\nonumber\\&-&
{1\over 2}\sum_{k=o}^{N-3}{x^{N-k-1}-y^{N-k-1}\over x-y} {\partial {\cal
U}_{N-k-1}\over \partial q_{n+1}}.\label{pp}
\end{eqnarray}
\end{prop}

\noindent
{\bf Proof.}  Let us rewrite the relation (\ref{rsal1}) in extended form

\begin{eqnarray}
\{U(x),U(y)\}&=&\{V(x),V(y)\}=0,\label{uu}\\
\{W_N(x),W_N(y)\}&=&4\alpha_N(x,y)(V(x)-V(y)),\label{ww}\\
\{V(x),U(y)\}&=&{2\over y-x}(U(x)-U(y)),\label{vu}\\
\{W_N(x),U(y)\}&=&{4\over x-y}(V(x)-V(y)),\label{wu}\\
\{V(x),W_N(y)\}&=&{2\over x-y}(W_N(x)-W_N(y))-2\alpha_N(x,y)U(x)
\label{vw}
\end{eqnarray}
with respect to the standard Poisson bracket. The equalities (\ref{uu},\ref{vu},\ref{wu})
can be proved directly from the definitions
(\ref{uz},\ref{vz},\ref{wz}). Let us prove (\ref{ww}) by  induction
i.e. let (\ref{ww}) be valid for the number $N$. Then for $N+1$ we have
\begin{eqnarray}
&&\{W_{N+1}(x),W_{N+1}(y)\}=\{W_{N+1}^+(x),W^-(y)\}
-\{W_{N+1}^+(y),W^-(x)\}\nonumber\\
&&=\{xW_{N}^+(x)-2{\cal U}_N,W^-(y)\}
-\{yW_{N}^+(y)-2{\cal U}_N,W^-(x)\}\nonumber\\
&&=x(\{W_{N}^+(x),W^-(y)\}-\{W_{N}^+(y),W^-(x)\})\nonumber\\&&-y(\{W_{N}^+(y),
W^-(x)\}-\{W_{N}^+(x),W^-(y)\})\nonumber\\
&&=x\{W_{N}^+(y),W^-(x)\}-y\{W_{N}^+(x),W^-(y)\}-2\{{\cal
U}_{N},W^-(y)\}\label{1}
\end{eqnarray}
the inductive hypothesis gives
\begin{equation}
\{W_{N+1}(x),W_{N+1}(y)\}-4\alpha_{N+1}(x,y)(V(x)-V(y))=\Delta\label{2}
\end{equation}
with
\begin{eqnarray}
\Delta & = & 2\sum_{i=1}^n {1\over{(x+A_i)(y+A_i)}} \left({2p_iq_i}(xQ_N(y)-yQ_N(x))\right.\nonumber\\
&+&\left. 2(x-y){\partial {\cal U}_N\over \partial q_i}\right.\nonumber\\
 &-&\left. {p_i} \left( x(y+A_i){\partial W^+_N(y)\over
\partial q_i}-y(x+A_i){\partial W^+_N(x)\over \partial
q_i} \right) \right)\nonumber\\
&=&\sum_{i=1}^n{\Delta_i}.\label{3a}
\end{eqnarray}
To prove that $\Delta=0$ we
substitute the expressions (\ref{sumq}), (\ref{sumw}) into $\Delta_i$. We
have
\begin{eqnarray}
\Delta_i&=&4q_i\left(xy^{N-2}-yx^{N-2}-{1\over
2}\sum_{k=0}^{N-3}(xy^k-yx^k){\partial {\cal U}_{N-k-1}\over \partial
q_{n+1}}\right)\nonumber\\
&+&4\sum_{k=0}^{N-2}\left(x(y+A_i)y^k-y(x+A_i)x^k\right){\partial {\cal
U}_{N-k-1}\over \partial q_i}\nonumber\\
&+&4(x-y){\partial {\cal U}_{N}\over \partial q_{i}}.
\end{eqnarray}
>From (\ref{equ1}) it follows that all $\Delta_i=0$ and the derivation of
(\ref{ww}) is complete.

Eq. (\ref{vw}) we also prove by induction. Assume that it is valid at
$N$. Then
\begin{eqnarray}
\{V(x),W_{N+1}(y)\}&=&
y\{V(x),W_N(y)\}+(1-y)\{V(x),W^-(y)\}
\nonumber\\&-&2\{V(x),{\cal U}_N\}\nonumber\\
&=&{2y\over x-y}(W_N^+(x)-W_N^+(y))+{2y\over x-y}(W^-(x)-W^-(y))\nonumber\\
&-&2yU(x)\alpha_N(x,y)+(1-y)\{V(x),W^-(y)\}\nonumber\\&-&2\{V(x),{\cal
U}_N\}\nonumber
\end{eqnarray}
Using the inductive proposition and the formulae
(\ref{w+-},\ref{w+},\ref{w-}) we find
\begin{eqnarray}
\{V(x),W_{N+1}(y)\}&-&{2\over y-x}(W_N(x)-W_N(y))+2\alpha_{N+1}(x,y)U(x)=\tilde
\Delta\nonumber
\end{eqnarray}
with
\begin{eqnarray}
\tilde\Delta&=&{2(y-1)\over
x-y}(W^-(x)-W^-(y))+(1-y)\{V(x),W^-(y)\}\nonumber\\
&-&2W^++2U(x)Q_N(x)-2\{V(x),{\cal U}_N\}\label{7}
\end{eqnarray}
The first two
terms in (\ref{7}) cancel due to the definitions (\ref{vz},\ref{w-}). To
cancel the rest we compute the Poisson bracket directly and use (\ref{wz}).
Therefore (\ref{vw}) is proven.

The equality (\ref{rsal1}) contains all the information concerning the
hierarchy of dynamical system under consideration.

Let us write the relation (\ref{rsal1}) in the form
\begin{equation}
\{L_1^{(N)}(x)\PB L_2^{(N)}(y)\}  =
[d^{(N)}_{12}(x,y),L_1^{(N)}(x)]-[d^{(N)}_{21}(x,y)L_2^{(N)}(y)],
\label{rsal2}
\end{equation}
 with $d^{(N)}_{ij}=r_{ij}+s_{ij},\,d^{(N)}_{ji}=s_{ij}-r_{ij}$ at $i<j$.

Then \cite{bv90}
\begin{equation}
\{(L_1^{(N)}(x))^k\PB(L_2^{(N)}(y))^l\}  =
[d^{(N,k,l)}_{12}(x,y),L_1^{(N)}(x)]-[d^{(N,k,l)}_{21}(x,y)L
_2^{(N)}(y)],
\label{rsal3}
\end{equation}
with
\begin{eqnarray}
&&d^{(N,k,l)}_{12}(x,y)=\label{dd}\\
&&\quad\sum_{p=0}^{k-1}\sum_{q=0}^{l-1}(L_1^{(N)}(x))^{k-p-1}
(L_2^{(N)}(y))^{k-q-1}d^{(N)}_{12}(x,y)(L_1^{(N)}(x))^{p}(L_2^{(N)}(y))^{q},\nonumber\\
&&d^{(N,k,l)}_{21}(x,y)=\nonumber\\
&&\quad\sum_{p=0}^{k-1}\sum_{q=0}^{l-1}(L_1^{(N)}(x))^{k-p-1}
(L_2^{(N)}(y))^{k-q-1}d^{(N)}_{21}(x,y)(L_1^{(N)}(x))^{p}(L_2^{(N)}(y))^{q}.\nonumber
\end{eqnarray}

As an immediate consequence of (\ref{rsal3}) and (\ref{dd}) we obtain the proof that the
conserved quantities $H_N$, $F_N^{(i)}$ are in involution.  We have that
\begin{equation}
\{{\rm Tr}(L_1^{(N)}(x))^2,{\rm Tr}(L_2^{(N)}(y))^2\}=
{\rm  Tr}\{(L_1^{(N)}(x))^2 \PB (L_2^{(N)}(y))^2\}\end{equation}
Applying   to   the   last
equation the equality (\ref{rsal3}) at $k=l=2$ and taking the trace we obtain  the
desired involutivity.

The Lax representation (\ref{*}) also can be recovered from
 (\ref{rsal1}).  We have
\begin{equation}
\dot L^{(N)}(x)=\frac18\lim_{y\rightarrow
\infty}{\rm Tr}_2\{L_1^{(N)}(x)\PB(L_2^{(N)}(y))^2\},\label{lax}
\end{equation}
where the trace is taken over the second space.  Applying the identity
(\ref{rsal3}) at $k=1,l=2$ to (\ref{lax}) we obtain the
Lax representation  $\dot L^{(N)}(x)=[M^{(N)}(x), L^{(N)}(x)]$
 with  the
 matrix  $M^{(N)}(x)$ given as
\begin{equation}
M^{(N)}(x)=2{\rm lim}_{y\rightarrow \infty} {\rm Tr}_2
L_1^{(N)}(y)(r(x-y)-s^{(N)}(x,y)).
\end{equation}
After the calculation in  which we take into
the  account  the  asymptotic
assumptions
\begin{eqnarray}
&&{\rm lim}_{y\rightarrow \infty}{U(y)\over y}= 4,\quad
{\rm lim}_{y\rightarrow \infty}{V(y)\over y}=0\nonumber\\
&&{\rm lim}_{y\rightarrow
\infty}\left(W_N(y)-U(y)Q_N(y)\right)=0\nonumber \end{eqnarray} we
obtain the Lax representation in the form (\ref{*}).  Analogously the
Lax representation for the higher flows can be obtained.

\subsection{Jacobi Identity}

To prove that the algebra (\ref{rsal}) is associative we write the Jacobi
identity in the $V^2\otimes V^2\otimes V^2$ as
\begin{eqnarray}
&&\{L_1^{(N)}(x)\PB\{L_2^{(N)}(x)\PB L_3^{(N)}(y)\}\}+
\{L_3^{(N)}(z)\PB\{L_1^{(N)}(x)\PB L_2^{(N)}(y)\}\}\nonumber\\
&&\quad+\{L_2^{(N)}(y)\PB\{L_3^{(N)}(z)\PB L_1^{(N)}(x)\}\}=0\label{Jaid}
\end{eqnarray}
with $L_1^{(N)}(x)=L^{(N)}(x)\otimes I \otimes I$, $L_2^{(N)}(y)= I  \otimes L^{(N)}(x)\otimes
id$,
$L_3^{(N)}(z)= I \otimes I\otimes L^{(N)}(z)$.
Let us rewrite  (\ref{Jaid})  in  the form \cite{ma85},
\begin{eqnarray}
&&[L_1^{(N)}(x),[d^{(N)}_{12}(x,y),d^{(N)}_{13}(x,z)]+[d^{(N)}_{12}(x,y),d^{(N)}_{23}(y,z)]
+\nonumber\\
&&+[d^{(N)}_{32}(z,y),d^{(N)}_{13}(x,z)]]+\label{ybe3}\\
&&+[L_1^{(N)}(x), \{L_2^{(N)}(y)\PB d^{(N)}_{13}(x,z)\}-\{L_3^{(N)}(z)\PB
d^{(N)}_{12}(x,y)\}] +\nonumber\\
&&+\quad\mbox{ cyclic permutations}=0\nonumber
\end{eqnarray}
We shall show below, that in the cases considered the solutions
(\ref{ybe3}) satisfy  the three equations
\begin{eqnarray}
&&[d^{(N)}_{12}(x,y),d^{(N)}_{13}(x,z)]+[d^{(N)}_{12}(x,y),d^{(N)}_{23}(y,z)]
+[d^{(N)}_{32}(z,y),d^{(N)}_{13}(x,z)]+\nonumber\\
&&+\{L_2^{(N)}(y)\PB d^{(N)}_{13}(x,z)\}-\{L_3^{(N)}(z)\PB
d^{(N)}_{12}(x,y)\} +\nonumber\\
&&\quad+[c^{(N)}(x,y,z), L_2^{(N)}(y)-L_3^{(N)}(z)]=0\label{ybe4}
\end{eqnarray}
where  $c^{(N)}(x,y,z)$  is  some  matrix  dependent  on   dynamical
variables. The other two equations are obtained from (\ref{ybe4}) by
cyclic permutations.  We remark, that validity of the  equations
(\ref{ybe4}) with an arbitrary matrix $c^{(N)}(x,y,z)$ is sufficient for the
validity of (\ref{Jaid}) and therefore (\ref{ybe4}) can be interpreted
as some dynamical classical Yang--Baxter equations.  These  equations have an
extra term $[c,L^{(N)}_i-L^{(N)}_j]$ in comparison with the extended Yang--Baxter
equations in \cite{ma85}.

\begin{prop}
The following equality is valid for all the member of the hierarchy of
dynamical systems
\begin{eqnarray}
\{L_2^{(N)}(y)\PB s_{13}(x,z)\}& -&\{L_3^{(N)}(z)\PB s_{12}(x,y)\} =
2\beta_N(x,y,z)[P_{23},S_{13}+
S_{12}]\nonumber\\&-&{\partial \beta_N(x,y,z)\over
\partial q_{n+1}}[s,L_2^{(N)}(y)-L_3^{(N)}(z)]\label{ybe1}
\end{eqnarray}
with  cyclic permutations.  In (\ref{ybe1}) the matrix $s=\sigma_-\otimes
\sigma_-\otimes \sigma_-$ and
\begin{equation}
\beta_N(x,y,z)={Q_N(x)(y-z)+Q_N(y)(z-x)+Q_N(z)(x-y)\over
(x-y)(y-z)(z-x)}\label{hh}
\end{equation}
\end{prop}

\noindent
{\bf Proof.}  Let us write the equality (\ref{ybe1}) in the extended
form
\begin{eqnarray}
\{Q(x),Q(y)\}&=&\{U(x),Q(y)\}=0\label{qq}\\
\{V(x),Q(y)\}&=&4\alpha_N(x,y)-{1\over 2}{\partial \alpha_N(x,y)\over \partial
q_{n+1}}U(x),\label{vq}\\
\{Q(x),W(y)\}&+&\{W(x),Q(y)\}={\partial \alpha_N(x,y)\over \partial
q_{n+1}}(V(x)-V(y)),\label{wq}\\
\{V(z),\alpha_N(x,y)\}&=&{4\over x-y}(\alpha_N(z,x)-\alpha_N(z,y))\nonumber\\
&-&{1\over 2} {U(z)\over x-y}\left({\partial \alpha_N(z,y)\over \partial
q_{n+1}}-{\partial \alpha_N(z,x)\over \partial
q_{n+1}}\right),\label{wp}
\end{eqnarray}
(\ref{qq}) is trivial and (\ref{vq}) and (\ref{wq}) follows immediately from
(\ref{vw}), (\ref{ww}) and (\ref{qw}). To prove (\ref{wp}) we write it using
the properties of $W_N(x)$ and $V(x)$ in the form
\begin{eqnarray}
&&\sum_{i=1}^n{p_i\over (y+A_i)(x+A_i)}\left((x+A_i){\partial \alpha_N(z,x)\over
\partial q_i}-(y+A_i){\partial \alpha_N(z,y)
\over \partial q_i}\right)\nonumber\\
&&={1\over 2}\sum_{i=1}^n{p_iq_i\over (x+A_i)(y+A_i)}
\left({\partial \alpha_N(z,x)\over \partial q_{n+1}}-{\partial
 \alpha_N(z,y)\over \partial q_{n+1}}\right)\label{10}
\end{eqnarray}
Using the decomposition (\ref{pq}), the equality
\[
x{z^k-x^k\over z-x}-y{z^k-y^k\over z-y}={z^{k+1}-x^{k+1}\over
z-x}-{z^{k+1}-y^{k+1}\over z-y}
\]
and the recurrence relation (\ref{equ1}) we find that the equality (\ref{10})
is valid.

Note that (\ref{vq}) can we rewritten in the
form
\begin{eqnarray}&&\{W_N(y),\alpha_N(z,x)\}-\{W_N(x),\alpha_N(z,y)\}\nonumber\\&&={V(
x)-V(y)\over
x-y} \left({\partial \alpha_N(z,y)\over \partial q_{n+1}}-{\partial
\alpha_N(z,x)\over \partial q_{n+1}}\right),\label{pq}
\end{eqnarray}

All the preceding material can be formulated in the following theorem

\begin{theorem}
Let
\begin{equation}
\{L_1^{(N)}(x)\PB L_2^{(N)}(y)\}=[d^{(N)}_{12}(x,y),L_1^{(N)}(x)]-
                              [d^{(N)}_{21}(x,y),L_2^{(N)}(y)]
\label{alg4}
\end{equation}
be the $r$-matrix algebra
constrained by the three conditions on the matrices $d^{(N)}_{ij}$  the
first  of which is written as
\begin{eqnarray}
&&[d^{(N)}_{12}(x,y),d^{(N)}_{13}(x,z)]+[d^{(N)}_{12}(x,y),d^{(N)}_{23}(y,z)]
+[d^{(N)}_{32}(z,y),d^{(N)}_{13}(x,z)]+\nonumber\\
&&+\{L_2^{(N)}(y),d^{(N)}_{13}(x,z)\}-\{L_3^{(N)}(z)\PB
d^{(N)}_{12}(x,y)\} +\nonumber\\
&&\quad+[c^{(N)}(x,y,z)\PB L_2^{(N)}(y)-L_3^{(N)}(z)]=0\label{ybe5}
\end{eqnarray}
and the two other are obtained  from  (\ref{ybe5})  by
cyclic permutations. Then there exists a  solution of
the equations  (\ref{ybe5})  describing  the  dynamics  of  the
hierarchy of completely integrable systems in the $n+1$ dimensional phase
space $(p_1,q_1,\ldots,p_{n+1},q_{n+1})$ given by the formulae
\begin{eqnarray}
d^{(N)}_{ij}(x,y)&=&{P_{ij}\over x-y}+{Q_N(x)-Q_N(y)\over x-y}S_{ij},\,\\
d^{(N)}_{ji}(x,y)&=&-{P_{ij}\over  x-y}+{Q_N(x)-Q_N(y)\over x-y}S_{ij}, \,
i<j,\,i,j=1,2,3,\nonumber\\
c^{(N)}(x,y,z)&=&{\partial \over \partial q_{n+1}}
{Q_N(x)(y-z)+Q_N(y)(z-x)+Q_N(z)(x-y)\over
(x-y)(y-z)(z-x)}s\nonumber
\end{eqnarray}
with
\begin{equation}
Q_N(z)=z^{N-2}-{1\over 2}\sum_{k=0}^{N-3}{\partial {\cal U}_{N-k-1}\over
\partial q_{n+1}} z^k
\end{equation}
and the
potentials  ${\cal  U}_N$  defined  as   (\ref{genpot}).  The associated   Lax
representation has the form (\ref{*})  with the matrix elements given by (\ref{uz},\ref{vz},\ref{wz}).
\end{theorem}

\subsection{Embedding into a Larger System}

In conclusion we  point out another direction in the algebraic
interpretation of the  hierarchy of dynamical systems considered. The
hierarchy can be embedded into a system with more degrees of freedom,
which admits the standard linear $r$-matrix algebra ($s=0$),
\begin{equation}
\{L_1^{(N)}(x)\PB L_2^{(N)}(y)\}=[r(x-y),L_1^{(N)}(x)+L_2^{(N)}(y)].\label{sal}
\end{equation}

To describe this embedding we introduce the following ansatz for the
functions $ U(z),\,V(z),\,W(z)$
\begin{eqnarray}
U(z)&=&\sum_{i=0}^{N-2}U_iz^i-\sum_{i=1}^n{q_i^2\over z+A_i},
\label{ansu}\\
V(z)&=&\sum_{i=0}^{N-2}V_iz^i+\sum_{i=1}^n{q_ip_i\over z+A_i},
\label{ansv}\\
W_N(z)&=&4z^{N-1}+\sum_{i=0}^{N-2}W_i^{(N)}z^i+\sum_{i=1}^n{p_i^2\over
z+A_i},\label{answ}
\end{eqnarray}
where the coefficients $U_i,V_i,W_i^{(N)}$ depends on the variables
$q_{n+1},p_{n+1}$ and in addition $N-2$ new canonically conjugated variables ${\cal Q}_j,{\cal P}_j$, $j= 1,\ldots N-2$. The substitution of the ansatz
(\ref{ansu})--(\ref{answ}) into (\ref{sal}) leads to the Lie algebra
\begin{eqnarray}
\{U_i,U_j\}&=&\{V_i,V_j\}=\{W_i^{(N)},W_j^{(N)}\}=0,\,i,j = 0,\ldots,N-2,
\nonumber\\
\{U_l,V_k\}&=&-2U_{l+k+1}\theta(k+l-N-1),\nonumber\\
\{V_l,W_k^{(N)}\}&=&2W_{l+k+1}\theta(k+l-N-1),\nonumber\\
\{W_l^{(N)},U_k\}&=&4U_{l+k+1}\theta(k+l-N-1), k,l=0,\ldots,N-2,\nonumber\\
\{V_i,W_{N-i-2}^{(N)}\}&=&8,\,i=0,\ldots,N-2,\label{lie}
\end{eqnarray}
where $\theta(n)=1$ if $n>0$ and $\theta(n)=0$ if $n\leq 0$.
The representations of this algebra give the algebraic description of the
 hierarchy of integrable systems considered.

Let us discuss, as an example, the case $N=3$ - many-particle
H\'enon-Heiles system.  We have only the coefficients
$U_0,U_1,V_0,V_1,W_0^{(3)},W_1^{(3)}$ in the ansatz (\ref{ansu})-(\ref{answ}) and
the Lie algebra for them is
\begin{eqnarray}
\{U_0,U_1\}&=&\{V_0,V_1\}=\{W_0^{(3)},W_1^{(3)}\}=0,\nonumber\\
\{U_1,V_1\}&=&\{U_1,W_1^{(3)}\}=\{V_1,W_1^{(3)}\}=0,\nonumber\\
\{V_0,U_0\}&=&-2U_1,\nonumber\\\{W_0^{(3)},U_0\}&=&4V_1,\nonumber\\\{V_0,W_0^{(3)}\}&=&2W_
1,\nonumber\\
\{V_1,W_0^{(3)}\}&=&\{V_0,W_1^{(3)}\}=8.\label{lie3}
\end{eqnarray}
The following representation for the algebra (\ref{lie3}) can be found
\begin{eqnarray}
U_0&=&4B-4q_{n+1}, \,U_1=4,\nonumber\\
V_0&=&2B{\cal P}_1+2p_{n+1}-2{\cal P}_1q_{n+1}, \,V_1=2{\cal P}_1,
\nonumber\\
W_0^{(3)}&=&4q_{n+1}^2+{\cal Q}_1-2{\cal P}_1p_{n+1}+{\cal P}_1^2q_{n+1},
\nonumber\\
W_1^{(3)}&=&4B+4q_{n+1}-{\cal P}_1^2..\label{uvw3}
\end{eqnarray}

The corresponding enlarged dynamical system admits the integrals
of motion ${\cal I}_3,\,{\cal H}_3, \,{\cal F}_3^{i}, i=1,\ldots,n$
given by the formulae
\begin{eqnarray}
{\cal I}_3&=&16{\cal Q}_1-4\sum_{m=1}^n q_m^2 +4B{\cal P}_1^2,
\label{intI}\\
{\cal H}_3&=&H_3+{1\over 8}\left[{\cal I}_3(B-q_{n+1})+{\cal P}_1
(4\sum_{i=1}^n P_i q_i +{\cal P}_1\sum_{i=1}^nq_i^2)\right]
\label{intH}\\
{\cal F}_3^{(i)}&=&F_3^{(i)}-{1\over 4}{\cal I}_3\sum_{i=1}^nq_i+
{\cal P}_1^2(B\sum_{i=1}^nq_i^2-{1\over
8}\sum_{i=1}^nA_iq_i^2-{1\over8}q_{n+1}\sum_{i=1}^nq_i^2)
\nonumber\\&&+{1\over2}{\cal
P}_1\left((B-q_{n+1})\sum_{i=1}^np_iq_i-\sum_{i=1}^nA_ip_iq_i
+{1\over4}p_{n+1}\sum_{i=1}q_i^2\right) \label{intF}
\end{eqnarray}
where in (\ref{intH}),(\ref{intF}) $H_3$ and $F_3^{(i)}$ are the integrals
of motions of the many-particle H\'enon-Heiles system calculated by
the formula (\ref{FN}).
We can see, that at ${\cal P}_1=0, {\cal I}_3=0$ the system is reduced
to the many-particle Henon--Heiles system.

The question of the explicit description of the other members of the
hierarchy within this approach will be given elsewhere.

\section{Separation of Variables}

\setcounter{equation}{0}
The separation of variables (c.f. \cite{sk89,ku92}) is
understood in the context of the given hierarchy of  Hamiltonian
system as the construction of $n+1$ pairs of canonical variables
$\pi_i,\,\mu_i,\,i=1,\ldots,n+1$,
\begin{equation}
\{\mu_i,\mu_k\}=\{\pi_i,\pi_k\}=0,\quad\{\pi_i,\mu_k\}=\delta_
{ik}\label{canvar}
\end{equation}
and $n+1$ functions $\Phi_j$ such that
\begin{equation}
\Phi_j\left(\mu_j,\pi_j,H_N,F_N^{(1)},\ldots,F_N^{(n)}\right)=0,\quad
j=1,2,\ldots,n+1,\label{Phi}
\end{equation}
where  $H_N,F_N^{(i)}$ are the integrals of motion in involution.
The equations (\ref{Phi}) are the separation
equations. The considered integrable system admits a Lax
representation in the form of $(2\times 2)$ matrices
(\ref{*}) and we will introduce the separation variables $\pi_i,\mu_i$ as
\begin{equation}
\pi_i=V(\mu_i),\quad U(\mu_i)=0,\quad
i=1,\ldots,n+1.\label{sepvar}
\end{equation}
Below we write these formulae explicitly for our system.

\subsection{Parabolic Coordinates}

The set of zeros $\mu_j,j=1,\ldots n+1$ of the  function  $U(z)$  in  the  Lax
representation (\ref{*})  defines  the  parabolic  coordinates  given  by  the
formulae \cite{kmw76,ku92}
\begin{eqnarray}
q_{n+1}&=&\sum_{i=1}^n A_i +B +\sum_{i=1}^{n+1} \mu_i,\nonumber\\ q_m^2
&=&-4{\prod_{j=1}^{n+1} (\mu_j+A_m)\over \prod_{k\neq m} (A_m-A_k)},\,
m=1,\ldots,n,\quad {\rm if} \quad n>1\label{q}
\end{eqnarray}
and
\begin{equation}q_1^2=-4(\mu_1+A)(\mu_2+A), \,A_1=A, \quad{\rm
if}\quad  n=1.
\end{equation}

Let us denote by  $\pi_m$,
\begin{equation}
\pi_m=V(\mu_i)=\dot \mu_m \prod_{i\neq m\atop i=1,\ldots,n+1}
{\mu_m-\mu_i\over \mu_m+A_i},\, m=1,\ldots,n+1.\label{pi}
\end{equation}
\begin{prop} The coordinates $\mu_i,\pi_i$ given by (\ref{q}),(\ref{pi})
are canonically conjugated.\end{prop}

{\bf Proof} Let us list the commutation relations betveen $U(z)$ and $V(z)$,
\begin{eqnarray}
\{U(x),U(y)\}&=&\{V(x),V(y)\}=0,\label{uuvv}\\
\{V(x),U(y)\}&=&{2\over y-x}(U(x)-U(y)),\label{vuvu}\\
\end{eqnarray}
The equalities $\{\mu_i,\mu_j\}=0,i,j=1,\ldots,n+1$  follows from
(\ref{uuvv}). To derive the equality
$\{\mu_i,\pi_j\}=\delta_{ij},i,j=1,\ldots,n+1$ we substitute $x=\mu_j$
in (\ref{vuvu}), obtaining thus \[ \{\pi_j,U(y)\}=-{2\over
y-\mu_j}U(y)),\] which together with the equation \[0=
\{\pi_j,U(\mu_i)\}=
\{\pi_j,U(y)\}\mid_{y=\mu_i}+U'(\mu_i)\{\pi_j,\mu_i\},\] gives

\[\{\pi_j,\mu_i\}=-{1\over
U'(\mu_i)}\{\pi_j,U(y)\}\mid_{y=\mu_i}=\delta_{ij}\] Equalities
$\{\pi_i,\pi_j\}=0,i,j=1,\ldots,n+1$ can be verified by the similar
way:  \begin{eqnarray} &&-\{\pi_i,\pi_j\}=
\{V(\mu_i),V(\mu_j)\}\nonumber\\
&&\{V(\mu_j),V(y)\}\mid_{y=\mu_i}+V'(\mu_i)\{\mu_i,V(\mu_j)\}\nonumber\\
&&V'(\mu_j)\{V(\mu_j),\mu_j\}\mid_{y=\mu_i}+V'(\mu_i)\{\mu_i,V(\mu_j)\}=0.
\nonumber\end{eqnarray}

The separation of variable equations have the form
\begin{equation}
\pi_i^2=w^2(\mu_i),\quad
i=1,\ldots,n+1,\label{sepeq}
\end{equation}
where  the  function  $w^2(z)$  is  given by (\ref{ccurve}).
Our use below of the separation equations is two-fold -- to integrate the
equations of motion in terms of theta functions and to quantize the
system.

\subsection{Theta Functional Integration of the Many Particle H\'enon--Heiles
System}

The curve (\ref{ccurve}) has genus $g=n+\left[N-1\over2\right]$. So only
in the cases $N=3,4$ does the number of degrees of freedom coincide with
the genus.  We consider here only the case $N=3$ (the Many-Particle
H\'enon-Heiles System) for which the curve (\ref{ccurve}) has
a branching point at infinity.  With this in mind we reduce
(\ref{sepeq})  to the Jacobi inversion problem
\begin{eqnarray}
&&\sum_{i=1}^{n+1}\int_{\mu_0}^{\mu_i} {\mu^k d\mu \over y(\mu)} =
C_k,\, k=0,\ldots,n-1,\nonumber\\&&\sum_{i=1}^{n+1}\int_{\mu_0}^{\mu_i}
{\mu^n d\mu \over y(\mu)} = it+C_n,\label{jac}
\end{eqnarray}
where $C_{n+1}=(y,z)$ is a nonsingular hyperelliptic curve of genus
$n+1$ given by the formula
\begin{eqnarray}
y^2= w^2 \prod_{m=1}^{n+1}
(z+A_m)=\prod_{k=1}^{2n+3}(z-z(Q_k)) ,\quad Q_i\neq Q_j
\end{eqnarray}
with the function $w^2$ given in (\ref{ccurve}) and $Q_1,\ldots$,
$Q_{2n+3}$,$Q_{2n+4}$, $z(Q_{2n+4})  =  \infty$ being the branching
points.

Let $({\bf a,b}) = (a_1,\ldots,a_{n+1};\,  b_1,\ldots,b_{n+1})$   be a
canonical base of  the first homology group  $H_1(C_{n+1},{\bf Z})$
with the intersection matrix,
\[
{\bf I}_{n+1} = \left(\begin{array}{cc}{\bf 0}_{n+1}&{\bf 1}_{n+1}\\
 -{\bf 1}_{n+1}&{\bf 0}_{n+1}\end{array}\right)\label{2.3}
\]
and let  ${\bf v} = ( v_1,\ldots,v_{n+1}),$
\begin{equation}
v_j=\sum_{k=1}^{n+1} c_{jk}w_{g-k+1}, \qquad w_k={z^{k-1}dz \over
y}, \qquad  j,k=1,\ldots,n+1\label{2.4}
\end{equation}
be the basis of holomorphic differentials normalized in such a way,
that the period matrix of $v_{n+1},\ldots,v_{n+1}$    with respect to
$({\bf a,b})$ has the form  $\Pi^T = ({\bf 1}_{n+1};{\tau })$,  where
${\bf 1}_{n+1}$    is the  $(n+1)\times (n+1)$ unit matrix, and the
$(n+1)\times (n+1)$   matrix  ${\tau}$
belongs to the Siegel upper half--space of the degree $ n+1$, ${\cal
S}_{n+1} = \{\tau \mid \tau  = \tau^T, {\rm Im} \tau > 0 \}$.
Let
${\bf D} = ({\cal A},{\cal B})$   be the divisor, where  ${\cal A}$ and
${\cal B}$     are positive divisors of degree  $g$;  let us denote the
Jacobian variety of the curve $C_{n+1}$ by  $J(C_{n+1}) = {\bf
C}^{n+1}/(1_{n+1},\tau)$   and the mapping ({\it Abel map}) which
establishes the correspondence between the point
${\bf e}=\left(\int_{\cal A}^{\cal B} {\bf v}\right) \in J(C_{n+1})$
and divisors  ${\cal D} \in C_{n+1}$    by $ {\cal D}  \Rightarrow
J(C_{n+1} )$.  We write ${\bf e}= {1\over 2}(\varepsilon',
\varepsilon'') \left[{}^{{\bf 1}_{n+1}}_{\tau}\right], \qquad
(\varepsilon', \varepsilon'') \in {\bf C}^{n+1}$, where $ [\varepsilon]
=\left[{}^{\varepsilon'}_{\varepsilon''}\right]$ is the  characteristic
of the point.  If  $\varepsilon_i', \varepsilon_j''$   is $0$ or
$1$,  the characteristics $[\varepsilon]$   are the characteristics of
half--periods.

We identify the branching points $Q_j,\, j=1,\ldots,2n+4 $ with the
characteristic of half-periods ${\bf e}_j \in J(C_{n+1})$ and fix for
definiteness the basis of homologies $({\bf a, b})$ as follows.
\begin{eqnarray}
{\bf e}_1  & =&\left[ {}^ {0  \ldots  0}_{ 0
               \ldots  0 }\right] , \quad {\bf e}_2
=\left[ {}_{ 1\,  0  \ldots  0}^{ 0\,  0  \ldots  0}\right] ,\quad {\bf
e}_3   =\left[{}_{ 1\, 0\,  0  \ldots  0}^{1\,  1 \, 0  \ldots
0}\right] , \ldots,\nonumber\\
{\bf e}_{ 2k+1 } &=& \left[ {}_{1\,  1  \ldots
1\,
0\,  0 \ldots  0}^{ 0\,  0  \ldots 0 \,1\, 0 \ldots 0} \right]
,\quad {\bf e}_{ 2k+2 } = \left[ {}_{ 1\,  1  \ldots  1
 1  \,0  \ldots  0}^{ 0\,  0  \ldots  0 1\, 0 \ldots 0 }\right] ,\ldots
, \nonumber\\{\bf e}_{ 2n+3 } &=&\left[ {}_{ 1\,  1  \ldots  1}^{1\,0\ldots 0}
\right] , \quad {\bf e}_{ 2n+4 }   =\left[ {}_{0\, 0 \ldots
0}^{1\,0\ldots 0} \right]. \label{2.5}
\end{eqnarray}
It is naturally to chose the ${\bf a}$-cycles to be homological to the
trajectories of $\mu$ variables under the evolution of the system .
The vector of Riemann constant ${\bf K}$  in the basis (\ref{2.5})
is ${\bf K}= \left[{}_{n+1\,n\ldots1}^{n+1\,1\ldots1}\right]$

The Riemann theta function $\Theta [\varepsilon]({\bf z}|\tau)$ with
the   characteristic $[\varepsilon]   = \left[ {}^{ \varepsilon'
}_{\varepsilon''}\right]$ is determined on  ${\bf C}^{n+1} \times {\cal
S}_{n+1}$ as
\begin{eqnarray}
\Theta [\varepsilon]({\bf
z}|\tau)&=&\sum_{{\bf m} \in {\bf Z}^{n+1}} \exp \pi \sqrt
{-1}\{\langle({\bf m} + {\varepsilon'\over2})\tau, ({\bf m}
+{\varepsilon'\over2}) \rangle
 \nonumber\\&+& 2\langle ({\bf m} + {\varepsilon'\over2}), {\bf z} +
{\varepsilon''\over2} \rangle\},\label{2.7}
\end{eqnarray}
where
$\langle \cdot , \cdot\rangle$ means the Euclidean scalar product.
Let us consider the curve with the ordering of the branching points
\begin{equation}
Q_{2k}=A_k,\quad k=1,\ldots,n.
\end{equation}
Using known hyperelliptic theta formulae \cite{fa73},
\begin{eqnarray}
\prod_{j=1}^{n+1}(\mu_j+A_k)=h_k{\Theta(\int_{Q_0}^{Q_k} {\bf v}+
(\int_{Q_0}^{\mu_1}+\ldots+\int_{Q_0}^{\mu_{n+1}}){\bf v}+{\bf K};\tau)
\over\Theta(\int_{Q_0}^{Q_{2n+4}}{\bf v} +
(\int_{Q_0}^{\mu_1}+\ldots+\int_{Q_0}^{\mu_{n+1}}){\bf v}+{\bf
K};\tau)},
\end{eqnarray}
where $k=1,\ldots,n$ and  $h_k$ are the
constants we obtain
\begin{eqnarray}
q_k^2(t)&=&q_k^2(0){\Theta^2[{\bf
e}_{2n+4}+{\bf K}](\omega_0;\tau)\Theta^2[{\bf e}_{2k+2} +{\bf
K}](\omega t+\omega_0;\tau)\over \Theta^2[{\bf e}_{2k+2}+{\bf
K}](\omega_0;\tau) \Theta^2[{\bf e}_{2n+4}+{\bf K}](\omega
t+\omega_0;\tau)},\nonumber\\
q_{n+1}(t)&=&B+\sum_{j=1}^n
A_j+\sum_{i=1}^{n+1}\oint_{a_i}z{v_i}(z)\nonumber\\&&-{\partial^2\over
\partial t^2} {\rm ln}\Theta[{\bf e}_{2n+4}+{\bf K}](\omega
t+\omega_0;\tau),
\end{eqnarray}
where the ``winding vector" $\omega$ is
expressed in terms of ${\bf b} $-periods of the differential of the
second kind $\Omega$ with zero ${\bf a}$--periods
and a unique pole at infinity which is expanded as $\Omega(z)=
{d\zeta / \zeta^2}+o(1)d\zeta,\, z=1/\zeta^2=\infty$ as
\[
\omega = {1\over 16
\pi i}\left(\oint_{b_1}\Omega,\ldots,\oint_{b_{n+1}}\Omega\right)
\]
and  $\omega_0\in {\bf C}^{g+1}$ is a constant.

The set of angles $\theta_i=\omega_it+\omega_{i0},\,i=1,\ldots,n+1$
 and the $n+1$ ${\bf a}$-periods of Abelian differentials
\begin{equation}
J_k=\oint_{a_k} \pi_k d\mu_k = \oint_{a_k} w
dz,\,k=1,\ldots,n+1,\label{action}
\end{equation}
constitute the ``action-angle" variables $J_k,\theta_k$ for the
system. The standard equality
\begin{equation}
H_3=\sum_{i=1}^{n+1} \omega_iJ_i
\end{equation}
follows from the Riemann bilinear identity.

The theta functional formulae for the evolution of the system can
also be given for the case $N=4$, where the curve (\ref{ccurve})
has no branching points at infinity, but its genus coincides with
the number of degrees of freedom (see, e.g. \cite{bbeim94}).
The algebro-geometric integration of the systems corresponding
to the higher members of hierarchy $(N>4)$ requires special
consideration.

\subsection{Quantization}
The separation of variables has a direct quantum counterpart
\cite{sk92a,kuz92}. To pass to quantum mechanics
 we change the  variables  $\pi_i,\mu_i$  to
operators and the Poisson brackets (\ref{canvar}) to the commutators
\begin{equation}
[\mu_i,\mu_k]=[\pi_i,\pi_k]=0,\quad[\pi_i,\mu_k]=-i\delta_{ik}\label{qcanvar}
\end{equation}
Suppose that the common spectrum of $\mu_i$ is simple and the momenta $\pi_i$
are realized as the differentials
$\pi_j=-i{\partial\over \partial \mu_j}$.    The separation equations
(\ref{sepeq})  become the operator equations,  where the noncommuting
operators are assumed to  be  ordered  precisely in the  order as those
listed in (\ref{Phi}), that is
$\pi_i,\mu_i,H_N,F_N^{(1)}$, $\ldots,F_N^{(n)}$.
Let $\Psi(\mu_1,\ldots,\mu_{n+1})$ be a common eigenfunction  of  the  quantum
integrals of motion:
\begin{equation}
H_N\Psi=\lambda_{n+1}\Psi,\,F_N^{(i)}\Psi=\lambda_i\Psi, i=1,\ldots,n.
\end{equation}
Then the operator separation equations lead to the set
of differential equations
\begin{equation}
\Phi_j(-i{\partial \over \partial
\mu_i},\mu_i,H_N,F_N^{(1)},\ldots,F_N^{(n)})\Psi(\mu_1,\ldots,\mu_{n+1})=0,\quad
j=1,\ldots,n+1,\label{oper}
\end{equation}
which allows the separation of variables
\begin{equation}
\Psi(\mu_1,\ldots,\mu_{n+1})=\prod_{j=1}^{n+1}\psi_j(\mu_j).
\end{equation}
The original multidimensional spectral problem is therefore reduced to the set
of one-dimensional multiparametric spectral problems which have the  following
form in the context of the  problem under consideration
\begin{equation}
\left({d^2\over d \,x^2}+16x^{N-2}(x+B)^2
+8\lambda_{n+1}+\sum_{i=2}^{n}{\lambda_i\over x+A_i}\right)
\psi_j(x;\lambda_1,\ldots,\lambda_{n+1})=0\label{quan}
\end{equation}
with the spectral parameters $\lambda_1,\ldots,\lambda_{n+1}$
The problem (\ref{quan}) must be solved on  the $n+1$
different intervals -- ``permitted zones":
$x \in [Q_1,Q_2],\ldots,$
$ [Q_{2n+1},Q_{2n+2}]$  for  the  variable  $x$.

\section*{Acknowledgements}
The authors are grateful to P P Kulish and E K Sklyanin
and M A Semenov-Tyan-Shanskii for  valuable discussions. We also
would like to acknowledge the EC for funding under the Science
programme SCI-0229-C89-100079/JU1.  One of us (JCE) is grateful to the
NATO Special Programme Panel on Chaos, Order and Patterns for support
for a collaborative programme, and to the SERC for research funding
under the Nonlinear System Initiative.

\end{document}